\documentclass[journal,twoside,web]{ieeecolor}
\usepackage{tmi}
\usepackage{cite}
\usepackage{amsmath,amssymb,amsfonts}
\usepackage{algorithm,algorithmic}
\usepackage{graphicx}
\usepackage{textcomp}
\usepackage{mathrsfs}
\def\BibTeX{{\rm B\kern-.05em{\sc i\kern-.025em b}\kern-.08em
    T\kern-.1667em\lower.7ex\hbox{E}\kern-.125emX}}
\begin{document}
\title{Uncertainty Propagation from Projections to Region Counts in Tomographic Imaging: Application to Radiopharmaceutical Dosimetry}
\author{Lucas Polson, Sara Kurkowska, Chenguang Li, Pedro Esquinas, Peyman Sheikhzadeh, Mehrshad Abbasi, Francois Benard, Carlos Uribe, Arman Rahmim
\thanks{Manuscript received XX; accepted XX. Date of publication XX; date of current version XX. This work was supported by the Natural Sciences and Engineering Research Council of Canada (NSERC) CGS D Award 569711, NSERC Discovery Grants RGPIN-2019-06467 and RGPIN-2021-02965, as well as computational resources and services provided by Microsoft for Health. \textit{(Carlos Uribe and Arman Rahmim contributed equally to this work.)}}
\thanks{Lucas Polson, Chenguang Li, and Arman Rahmim are with the the Department of Physics\& Astronomy at the University of British Columbia, Vancouver, Canada (e-mail: lukepolson@outlook.com).}
\thanks{Sara Kurkowska, Pedro Esquinas, Francois Benard, and Carlos Uribe are with the Department of Integrative Oncology, BC Cancer, Vancouver Canada.}
\thanks{Peyman Sh.Zadeh and Mehrshad Abbasi are with the Nuclear Medicine Department, IKHC, Faculty of Medicine, Tehran University of
Medical Science, Tehran, Iran}}

\maketitle

\begin{abstract}
Radiopharmaceutical therapies (RPTs) present a major opportunity to improve cancer therapy. Although many current RPTs use the same injected activity for all patients, there is interest in using absorbed dose measurements to enable personalized prescriptions. Image-based absorbed dose calculations incur uncertainties from calibration factors, partial volume effects and segmentation methods. While previously published dose estimation protocols incorporate these uncertainties, they do not account for uncertainty originating from the reconstruction process itself with the propagation of Poisson noise from projection data. This effect should be accounted for to adequately estimate the total uncertainty in absorbed dose estimates. This paper proposes a computationally practical algorithm that propagates uncertainty from projection data through clinical reconstruction algorithms to obtain uncertainties on the total measured counts within volumes of interest (VOIs). The algorithm is first validated on ${}^{177}$Lu and ${}^{225}$Ac phantom data by comparing estimated uncertainties from individual SPECT acquisitions to empirical estimates obtained from multiple acquisitions. It is then applied to (i) Monte Carlo and (ii) multi-time point ${}^{177}$Lu-DOTATATE and ${}^{225}$Ac-PSMA-617 patient data for time integrated activity (TIA) uncertainty estimation. The outcomes of this work are two-fold: (i) the proposed uncertainty algorithm is validated, and (ii) a blueprint is established for how the algorithm can be inform dosimetry protocols via TIA uncertainty estimation. The proposed algorithm is made publicly available in the open-source image reconstruction library PyTomography.
\end{abstract}


\section{Introduction}
\label{sec:introduction}
\IEEEPARstart{R}{adiopharmaceutical} therapies (RPTs) are rapidly advancing, with widespread use in clinical trials and recent approvals for the treatment of several cancers. ${}^{177}$Lu-based RPTs, for example, have achieved significant milestones in recent years. ${}^{177}$Lu-DOTATATE based peptide receptor RPT gained food and drug administration approval in 2018 for the treatment of neuroendocrine tumor patients \cite{lu177_dotatate}. Soon after, ${}^{177}$Lu-PSMA-617 was demonstrated efficacious in the treatment of metastatic castration resistant prostate cancer (mCRPC) in the VISION \cite{lu177_vision} and TheraP \cite{lu177_therap} randomized control trials. Numerous other clinical trials that focus on other isotopes have also shown positive results. Notably, recent studies on mCRPC RPTs have explored the use of ${}^{225}$Ac to enhance treatment outcomes\cite{Meyerjnumed.123.265433, kratochwil2016ac, ac225_meta}.

SPECT imaging can be used to monitor therapy in the aforementioned RPTs since photon emissions from direct ${}^{177}$Lu and indirect ${}^{225}$Ac decay are detectable by SPECT cameras. Imaging can be used to perform dosimetry after an initial therapeutic dose, with the goal of measuring absorbed dose in tumours and critical organs to personalize administered activity in later cycles of RPTs \cite{patient_specific}. When images are used to estimate absorbed dose, however, it may be important to consider the impact of limited statistical counts (LSCs) in projection data on the uncertainty in voxel values of the reconstructed images. While previous protocols have reported uncertainty analyses in RPT absorbed dose calculations \cite{Gear2018, nuttens_uncertainty, peters_uncertainty}, these protocols do not consider uncertainties propagating from LSCs in projection data.

In general, data obtained with any photon-based imaging modality are affected by some (perhaps negligible) amount of Poisson noise due to the finite number of photons detected. The noise is particularly significant in emission tomography due to the small number of photons detected, and it translates into corresponding noise and uncertainty in reconstructed images. Certain studies have aimed at quantifying these noise properties \cite{quan1,quan2}, while others have estimated noise via modification of filtered back projection (FBP) algorithms \cite{est1,est2}, or bootstrapping (FBP and iterative) \cite{est3}. Barrett et al.\ \cite{Barrett_1994} derived an iterative technique to estimate uncertainty propagation in the expectation maximization algorithm. Qi \cite{qi2003unified} later presented a more general formalism for uncertainty propagation in preconditioned gradient ascent (PGA) algorithms. 

These approaches by Barrett et al.\ and Qi, however, are computationally expensive to directly implement on clinical data due to large matrix-matrix products. The present work, as such, presents an extension of Qi's algorithm to make uncertainly estimation computationally efficient by considering the particular problem of total count uncertainty estimation within volumes of interest (VOIs). In what follows, the theoretical formalism is established, and the proposed algorithm is applied the specific and important task of absorbed dose uncertainty estimation in RPTs \cite{RPT}. To facilitate the integration of the proposed uncertainty estimation technique in such protocols, the algorithm was implemented in the open-source image reconstruction library PyTomography \cite{pytomo}.

\section{Methods}

This paper consists of three components: (i) derivation of the proposed uncertainty algorithm (ii) validation of the algorithm on physical ${}^{177}$Lu and ${}^{225}$Ac SPECT/CT phantom data, and (iii) application of the algorithm on multiple time point SPECT/CT imaging from ${}^{177}$Lu-DOTATATE and ${}^{225}$Ac-PSMA-617 RPTs, with further applications in time activity curve (TAC) fitting and time integrated activity (TIA) estimation. All reconstructions and uncertainty estimations in this paper were calculated with PyTomography \cite{pytomo}.

\subsection{Theory}

There are a few mathematical conventions used in this paper. The following notation is used for random variables: if $x$ is a random variable, then $\hat{x}$ corresponds to an analytically derived estimator for $x$ while $\tilde{x}$ corresponds to an empirical estimate of $x$ obtained by repetition of experiments. Uncertainties of an estimator are also estimators, and may be denoted by $\hat{u}(\hat{x})$, $\hat{u}(\tilde{x})$, $\tilde{u}(\hat{x})$, or $\tilde{u}(\tilde{x})$ depending on whether they are analytically or empirically obtained. The covariance matrix for a random variable $x$ consisting of more than one element is denoted $\Sigma_x$. The corresponding percent uncertainties are defined as $\delta(x) \equiv u(x)/x \cdot 100\%$, where the carets are omitted here. The following notation is used for vectors: $\cdot$ corresponds to the dot product, $\odot$ corresponds to the Hadamard or element-wise product, and $v^{\circ \alpha}$ corresponds to Hadamard or element-wise power of all elements in vector $v$ to the power $\alpha$. A primed linear operator $A'$ represents the transpose of linear operator $A$.

\subsubsection{VOI Uncertainty Estimation Formalism}

The random vector of the acquired data $y$ can be decomposed into an expectation value $\bar{y}$ and zero-mean noise vector $n$ as 

\begin{equation}
    y = \bar{y} + n \label{eq:y_noise}
\end{equation}
A reconstruction algorithm $A$ uses this projection data to obtain an image estimate $\hat{x}$ via

\begin{equation}
    \hat{x} = A(y) \label{eq:recon}
\end{equation}
Since $y$ is random vector, so too is $\hat{x}$; hence $\hat{x}$ can be written as

\begin{equation}
    \hat{x} = \bar{\hat{x}} + \epsilon \label{eq:x_noise}
\end{equation}
where $\epsilon$ is a zero-mean random noise vector. $\epsilon$ corresponds to an uncertainty resulting solely from uncertainty in the projection data. All information about $\epsilon$ is contained in $\Sigma_{\epsilon}$, which depends on $A$ and $\Sigma_n$. Provided $\Sigma_{\epsilon}$ can be estimated using some estimator $\hat{\Sigma}_{\epsilon}$, the uncertainty on the sum of voxel values within a volume of interest (VOI) mask $\xi$ can be estimated as

\begin{equation}
    \hat{u}(\hat{x} \cdot \xi) = \sqrt{\xi' \hat{\Sigma}_{\epsilon} \xi} \label{eq:var}
\end{equation}
 When expressed in the form of \ref{eq:var}, it is clear that $\hat{\Sigma}_{\epsilon}$ can be interpreted as an operator (much like the system matrix), and thus the individual components need not all be computed when operating on a specific VOI mask $\chi$. In what follows, the relationship between $\hat{\Sigma}_{\epsilon}$ and $\hat{\Sigma}_n$ is established for various reconstruction algorithms $A$.

\subsubsection{Covariance Matrix Estimation}

The most common linear reconstruction algorithm is filtered back projection (FBP). Using the distributive property of linear reconstruction algorithms and equation \ref{eq:recon}, the relationship between $\epsilon$ and $n$ can be shown to be given simply by

\begin{equation}
    \epsilon = A n
\end{equation}
implying

\begin{equation}
    \hat{\Sigma}_{\epsilon} = A \hat{\Sigma}_n A'
\end{equation}
Uncertainty estimation for linear reconstruction algorithms thus requires (i) an estimate for the covariance of the acquired data $\hat{\Sigma}_n$ and (ii) action of the adjoint $A'$ on $\xi$ in Eq.\ \ref{eq:var}. 

However, the majority of presently used tomographic reconstruction algorithms for image quantification are iterative in nature, and thus propagation of noise estimation is not straightforward. Barrett et al.\ \cite{Barrett_1994} derived an iterative technique to estimate uncertainty propagation in the EM algorithm. Meanwhile, many presently used iterative algorithms (e.g.\ EM) are actually a particular form of preconditioned gradient ascent (PGA) algorithms, and a framework for uncertainty propagation in PGA algorithms was in fact proposed by Qi \cite{qi2003unified}. PGA algorithms in image reconstruction have the following form:

\begin{align}
    \hat{x}^{k+1} = \hat{x}^{k} + & \text{diag}\left(\hat{x}^k\right) D(H) \bigl[ \nabla_{x} L\left( y|\hat{x}^{k},H,s \right)  \nonumber \\
    &- \beta \nabla_{x} U\left(\hat{x}^{k}\right) \bigr]
    \label{eq:pga}
\end{align}
where $\hat{x}^k$ is the image estimate of the $k$th iteration, $H$ is the system matrix for the imaging system, $D(H)$ is a positive definite matrix, $L(y|\hat{x}^{k},H)$ is the likelihood function characterizing the projection data $y$, $s$ is a an additive term used to account for scatter (in SPECT) and random/scatter (in PET), $U(\hat{x}^{k})$ is a regularization function, and $\beta$ is a scaling factor. Using the results from Qi \cite{qi2003unified} and including the estimated scatter uncertainty $\sigma$, it follows that
\begin{align}
    \epsilon^{k+1} \approx V_y^{k+1}n + V_s^{k+1}\sigma
    \label{eq:noise_iter}
\end{align}
where, letting $z$ be a placeholder for $y$ and $s$, it follows that
% &= \sum_{i=0}^{k-1} \left(\prod_{k-1}^{k-i} {Q_i} \right) B_z^{k-1-i} \nonumber \\
\begin{align}
    V_z^k &={B_z^{k-1}} + Q^{k-1} B_z^{k-2} + Q^{k-1}Q^{k-2} B_z^{k-3}  \nonumber \\ 
    & \hspace{6mm} + ... +   Q^{k-1} Q^{k-2} ... Q^2 Q^1 B_z^{0} 
    \label{eq:uncertainty_proj}
\end{align}
\begin{align}
    Q^k = 1 &+ \text{diag}\left(D^k\bar{\hat{x}}^k\right) \left[ \nabla_{xx} L(\bar{y}|\bar{\hat{x}}^k, \bar{s}) - \nabla_{xx} U(\bar{\hat{x}}^k)) \right] \nonumber\\
    &+ \text{diag} \left( D^k \left[\nabla_x L(\bar{y}|\bar{\hat{x}}^k, \bar{s}) -  \nabla_{x} U(\bar{\hat{x}}^k))\right] \right) \label{eq:Q}
\end{align}
\begin{equation}
    B_z^k = \text{diag}\left(\bar{\hat{x}}^k\right) D^k(H) \nabla_{xz} L(\bar{y}|\bar{\hat{x}}^k, \bar{s}) \label{eq:B}
\end{equation}

Using (\ref{eq:noise_iter}) and the fact that $n$ and $\sigma$ are independent, an estimator for the covariance matrix of $x^k$ can be derived as
\begin{align}
    \hat{\Sigma}_{\epsilon^k} = \hat{V}_y^k \hat{\Sigma}_n [\hat{V}_y^k]' + V_s^k \hat{\Sigma}_{\sigma} [\hat{V}_s^k]'
    \label{eq:cov_equation}
\end{align}
where $\hat{V}$ corresponds to substitution of $\hat{x}^k$, $y$, and $\hat{s}$ for $\bar{\hat{x}}^k$, $\bar{y}$, and $\bar{s}$ respectively in Eqs.\ \ref{eq:B} and \ref{eq:Q}. The form of $\hat{B}_z^k$ and $\hat{Q}^k$ in for the Poisson likelihood function are shown for ordered subset expectation maximization \cite{osem} and block sequential regularized expectation maximization \cite{bsrem} in Tab.\ \ref{tab:example}. For uncorrelated Poisson data, representative of what is measured in SPECT/PET, the covariance matrix of the photopeak is estimated by

\begin{equation}
    \hat{\Sigma}_n = \left\{ y \right\}
    \label{eq:sigma_n}
\end{equation}
In SPECT, the triple energy window method \cite{TEW} is used to estimate scatter via $\hat{s} = K_s[w_ls_l+w_us_u]$ where $w_l$ and $w_u$ are weightings for data acquired in different energy windows $s_l$ and $s_u$, and $K_s$ is a Gaussian smoothing kernel often used when scatter data is sparse. In this case

\begin{equation}
    \hat{\Sigma}_{\sigma} = K_s' \left\{w_l^2 s_l + w_u^2 s_u  \right\} K_s
    \label{eq:sigma_sigma}
\end{equation}

Substituting Eqs.\ \ref{eq:sigma_n}, \ref{eq:sigma_sigma}, and \ref{eq:post-filter} into \ref{eq:cov_equation} and combining with Eq.\ \ref{eq:var}, it follows that

\begin{align}
    \hat{u}(\hat{x}^k \cdot \xi) &= \biggl[ y \cdot \left([\hat{V}_y^k]' \xi\right)^{\circ 2} \nonumber \\
    &+ \left(w_l^2 s_l + w_u^2 s_u  \right) \cdot  \left(K_s[\hat{V}_s^k]' \xi\right)^{\circ 2} \biggr]^{1/2} \label{eq:final}
\end{align}
where $V_y$ and $V_s$ are specified by Eq.\ \ref{eq:uncertainty_proj}. For PET, where the scatter is estimated via single scatter simulation or Monte Carlo, and the resulting profile is smooth, the scatter term can be neglected. 

Often in low count scenarios, a post-reconstruction filter is applied to a reconstructed image to reduce noise. In general, if a linear operator $L_p$ is applied to a reconstructed image, then
\begin{align}
    &\epsilon \to L_p \epsilon \nonumber \\
    \implies & \Sigma_{\epsilon} \to L_p \Sigma_{\epsilon} L_p'
    \label{eq:post-filter}
\end{align}
Based on Eqs.\ \ref{eq:var} and \ref{eq:final}, consideration of this additional operation amounts to substitution of $\xi$ with $L_p' \xi$ in Eq.\ \ref{eq:final}. In this regard, post-reconstruction image filtering translates to pre uncertainty estimation mask filtering.

Eq.\ \ref{eq:final} is an extension of Qi's technique (Eq.\ \ref{eq:cov_equation}) that that reduces computation time by considering matrix-vector products instead of matrix-matrix products. While it yields less information than a complete covariance matrix between all voxel pairs, it yields information that is particularly well suited to estimate the uncertainty of time integrated activity (TIA) for dosimetry application.

\begin{table*}[h]
\centering
\begin{tabular}{l||l|l|l}
 Algorithm & $\hat{Q}^k$ & $\hat{B}_y^k$ & $\hat{B}_s^k$  \\
\hline
OSEM
& $\begin{aligned} 1 + \left\{\frac{\hat{x}^k}{H_k' 1}\right\}&\left[H_k' \left\{\frac{y_k}{(H_k\hat{x}^k+s)^2}\right\}  H_k \right]\\ &+ \hat{x}^k / \hat{x}^{k+1} \end{aligned}$  
& $\begin{aligned}
    \left\{\frac{\hat{x}^k}{H_k' 1}\right\} H_k' \left\{\frac{1}{H_k\hat{x}_k + s}\right\}
\end{aligned}$ 
& $\begin{aligned}
    -\left\{\frac{\hat{x}^k}{H_k' 1}\right\} H_k' \left\{\frac{y}{(H_k\hat{x}_k + s)^2}\right\}
\end{aligned}$\\
\hline
BSREM
& $\begin{aligned} 1 + \left\{\frac{|S_k|\hat{x}^k}{H' 1}\right\}&\left[H_k'  \left\{\frac{y_k}{(H_k\hat{x}^k+s)^2}\right\}  H_k \right.\\ & \left. - \nabla_{xx}U(\hat{x}^k) \right]+ \hat{x}^k / \hat{x}^{k+1} \end{aligned}$  
& $\begin{aligned}
    \left\{\frac{|S_k|\hat{x}^k}{H' 1}\right\} H_k' \left\{\frac{1}{H_k\hat{x}_k + s}\right\}
\end{aligned}$ 
& $\begin{aligned}
    -\left\{\frac{|S_k|\hat{x}^k}{H' 1}\right\} H_k' \left\{\frac{y}{(H_k\hat{x}_k + s)^2}\right\}
\end{aligned}$\\
\end{tabular}
\caption{SPECT reconstruction algorithms and corresponding operators required for computation of $V_z^k$ in Eq.\ \ref{eq:uncertainty_proj}. $|S_k|$ is the size of subset $k$ divided by the size of all subsets. For BSREM, a truncation is also typically applied after each iteration to ensure that all voxel values are greater than zero.}
\label{tab:example}
\end{table*}

\subsubsection{Time Integrated Activity}

This section demonstrates how the error estimation technique in the previous section can be used for dosimetry: in particular, how to determine uncertainties in the time integrated activity (TIA) for a specific VOI. The TIA in a VOI is defined as

\begin{equation}
    \mathscr{A} = \int_0^{\infty} f(t, \boldsymbol{p})dt
\end{equation}
where $f(t, \boldsymbol{p})$ is a function that describes the time activity curve. The TAC curve parameters $\boldsymbol{p}=[p_1,...,p_q]$ are estimated via fitting $n$ data points $(t_i, A_i)$ using non-linear regression techniques to minimize $\chi^2=\sum[(A_i-f(t_i, \boldsymbol{p}))/\sigma_i]^2$ where $t_i$ is the time, $A_i \equiv \hat{x}_i \cdot \xi$ is the voxel sum within a VOI, and $\sigma_i \equiv \hat{u}(\hat{x}_i \cdot \xi)$ is the estimated uncertainty. The estimated TIA is given by $\hat{\mathscr{A}} = g(\boldsymbol{\hat{p}})$, where $g$ is some function of the fit parameters. Since the estimated fit parameters $\hat{\boldsymbol{p}}$ depend on the random variables $A_i$ (with uncertainty characterized by $\sigma_i$), it follows that $\hat{\mathscr{A}}$ is itself a random variable; its true uncertainty can be defined as $u(\hat{\mathscr{A}})$. 

The estimated covariance matrix of the fit parameters $\hat{V}$ is assumed implicitly to be a function of the fit $\hat{\boldsymbol{p}}$, and can be used to obtain an estimate for $u(\hat{\mathscr{A}})$ via
\begin{equation}
    \hat{u}(\hat{\mathscr{A}}) = \left(\nabla_{\boldsymbol{p}}g\right)' \hat{V} \left(\nabla_{\boldsymbol{p}}g\right) \Big|_{\boldsymbol{p} = \hat{\boldsymbol{p}}} \label{eq:unc_cumulative_activity}
\end{equation}
The fit functions $f$, along with the corresponding $g$ and $\nabla_{\boldsymbol{p}} g$ used for various VOIs in this paper are shown in Tab.\ \ref{tab:fit_functions}.

\begin{table}[h]
\centering
\begin{tabular}{l||l|l}
 & Organs & Lesions \\
\hline
$f(t,\boldsymbol{p})$ & $p_0e^{-p_1 t}$ & $p_0(e^{-p_1 t} - e^{-p_2 t})$\\
$g(\boldsymbol{p})$ & $p_0/p_1$ & $p_0/p_1 - p_0/p_2$\\
$\nabla_{\boldsymbol{p}}g$ & $\left(p_1^{-1}, -p_0p_1^{-2}\right)$ & $\left(p_0^{-1},  -p_0p_1^{-2}, p_0p_2^{-2}\right)$\\
\end{tabular}
\caption{Activity fit functions $f(t,\boldsymbol{p})$ with corresponding TIA estimates $g(\boldsymbol{p})$ and gradients $\nabla_{\boldsymbol{p}}g$ used for various VOIs in this paper.}
\label{tab:fit_functions}
\end{table}

The provided values of $\sigma_i$ in curve fitting will impact the estimates $\hat{\boldsymbol{p}}$ and $\hat{V}$. There are three different ways to provide the $\sigma_i$ in curve fitting:

\begin{enumerate}
    \item Use of the estimated $\sigma_i$. In the context of TAC fitting, this involves using the uncertainties obtained from Eq.\ \ref{eq:final}. The corresponding curve fit estimates $\hat{\boldsymbol{p}}^{(\text{est-sig})}$ and $\hat{V}^{(\text{est-sig})}$ yield TIA and TIA uncertainty estimates of $\hat{\mathscr{A}}^{(\text{est-sig})}$ and $\hat{u}^{(\text{est-sig})}(\hat{\mathscr{A}}^{(\text{est-sig})})$.
    \item Use of ``proportional'' values $ \sigma_i / c$ in place of $\sigma_i$ where $c>0$ is an a priori unknown proportionality constant. For algorithms insensitive to scaling of $\chi^2$, $\hat{\boldsymbol{p}}^{(\text{prop-sig})} = \hat{\boldsymbol{p}}^{(\text{est-sig})}$, and the covariance matrix can be estimated as
    \begin{align}
     \hat{V}^{(\text{prop-sig})}(\left\{\sigma_i\right\}) &= \left( \frac{1}{\hat{c}^2} \right) \hat{V}^{(\text{est-sig})}(\left\{\sigma_i / c\right\}) \nonumber\\
     &=  \left(\frac{n-q}{\chi^2}\right) \hat{V}^{(\text{est-sig})}(\left\{\sigma_i / c\right\}) 
     \label{eq:rel2abs_V}
    \end{align}
    where $\hat{V}(\{\sigma_i / c\})$ results from use of $\sigma_i/c$ in a curve fitting procedure. These yield TIA and TIA uncertainty estimates of  $\hat{\mathscr{A}}^{(\text{prop-sig})}$ and $\hat{u}^{(\text{prop-sig})}(\hat{\mathscr{A}}^{(\text{prop-sig})})$
    \item Use $\sigma_i=1$ for all $i$. This is equivalent to providing equally weighted proportional uncertainties for each data point. The corresponding estimates $\hat{\boldsymbol{p}}^{(\text{no-sig})}$ and $\hat{V}^{(\text{no-sig})}$ yield TIA and TIA uncertainty estimates of $\hat{\mathscr{A}}^{(\text{no-sig})}$ and $\hat{u}^{(\text{no-sig})}(\hat{\mathscr{A}}^{(\text{no-sig})})$ respectively.
\end{enumerate}

It is demonstrated in the appendix that the count variance in mask $\xi_j$ at different counts is approximately given by

\begin{equation}
    \sigma_i \approx c_j \left(x_i \cdot \xi_j\right)^{1/2} \label{eq:prop_sigma}
\end{equation}
where $c_j$ is a proportionality factor for VOI $j$. In this case, $\hat{u}^{(\text{prop-sig})}(\hat{\mathscr{A}})$ can be obtained by using $\sigma_i / c= (x_i \cdot \xi)^{1/2}$ in Eq.\ \ref{eq:rel2abs_V}, and Eq.\ \ref{eq:final} is not needed. As will be shown in the subsequent sections, however, the reliance on $\hat{c}$ in Eq.\ \ref{eq:rel2abs_V} results in significantly reduced precision compared to using of the estimated uncertainties. Furthermore, Eq.\ \ref{eq:rel2abs_V} is unusable for any case where $n \leq q$ (e.g. a two time point fit to any curve with two or more parameters). In this scenario, estimation of $u(\hat{\mathscr{A}})$ requires $\sigma_i$ directly, so the uncertainties provided by Eq.\ \ref{eq:final} must be used.

\subsection{Validation and Clinical Application}

\subsubsection{Phantom-Based Validation of Uncertainty Estimation}
In what follows, the uncertainty estimation technique is validated for SPECT with NEMA phantom data reconstructed using two popular reconstruction protocols: (i) OSEM and (ii) BSREM with the relative difference penalty (RDP) \cite{RDP}. OSEM with one subset is denoted as maximum likelihood expectation maximum (MLEM). In particular, the estimated percent uncertainty, given by

\begin{equation}
    \hat{\delta}(\hat{x}^k \cdot \xi) = \frac{\hat{u}(\hat{x}^k \cdot \xi)}{\hat{x}^k \cdot \xi}
    \label{eq:rel_unc_est}
\end{equation}
is compared to the empirical uncertainty estimator obtained by taking multiple acquisitions in sequence:

\begin{equation}
    \tilde{\delta}(\hat{x}^k \cdot \xi) = \frac{\sqrt{\text{Var}\left[\left\{w_i \hat{x}^k_i \cdot \xi; 1 \leq i \leq N \right\}\right] }}{\text{Mean}(\left[ \left\{w_i \hat{x}^k_i \cdot \xi; 1 \leq i \leq N\right\} \right])}
    \label{eq:rel_unc_emp}
\end{equation}
where $i$ is the acquisition index, $N$ is the total number of acquisitions taken, and $w_i$ is an exponential scalar weighting factor used to compensate for activity decay over the course of multiple repeated acquisitions.

For ${}^{177}$Lu, a NEMA phantom with spheres of diameter $37~$mm, $28~$mm, 22~mm, 17~mm, 13~mm, and 10~mm was filled with a 9:1 source to background ratio with sphere activity concentration of $0.89~$MBq/mL. Fifty-three SPECT acquisitions of the phantom were taken in sequence on a Siemens Symbia T2 system with the following settings: $128 \times 128$ pixels at $4.82~\text{mm} \times 4.82~\text{mm}$ resolution, 96 projection angles, medium energy collimators, and 15~s acquisition time per projection. Acquired energy windows are shown in Tab.\ \ref{tab:energy_windows}. The data were reconstructed using (i) OSEM with 8 subsets and up to 10 iterations, and (ii) BSREM and the relative difference penalty (RDP) ($\beta=0.3$, $\gamma=2$) with 8 subsets and up to 40 iterations. Eight regions were considered for uncertainty estimation: the six spheres in the NEMA phantom, a background VOI consisting of two $50~$mm diameter spheres drawn in the warm region, and the central cold cylinder portion of the phantom. Sample reconstructions, estimated uncertainties, and true uncertainties for these use cases are shown in Fig.\ \ref{fig:lu177_nema}.

\begin{table}[h]
\centering
\begin{tabular}{l||c|c|c}
 \textbf{Experiment} & Photopeak & Lower Scatter\ & Upper Scatter\  \\
\hline
${}^{177}$Lu NEMA & 187.2, 228.8 & 166.4, 187.2 & 228.8, 249.6 \\
${}^{177}$Lu XCAT & 187.2, 228.8 & 169.4, 187.2 & 228.8, 252.9 \\
${}^{177}$Lu Patient & 187.6, 229.2 & 166.7, 187.6 & 229.2, 250.1 \\
${}^{225}$Ac Jaszczak & 196.2, 239.8 & 163.5, 196.2 & 239.8, 272.5 \\
 & 396.0, 484.0 & 352.0, 396.0 & - \\
${}^{225}$Ac XCAT & 196.2, 239.8 & 163.5, 196.2 & 239.8, 272.5 \\
 & 396.0, 484.0 & 352.0, 396.0 & - \\
${}^{225}$Ac Patient & 196.2, 239.8 & 177.5, 196.1 & 239.9, 265.1 \\
 & 396.0, 484.0 & 358.2, 395.9 & - \\
\end{tabular}
\caption{Acquired energy windows in units of keV. ${}^{225}$Ac contains both the 218~keV (${}^{221}$Fr) peak and the 440~keV (${}^{213}$Bi) peak. TEW scatter correction was used in cases where both the lower and upper windows were obtained, while dual energy window (DEW) was used when only the lower window was obtained.}
\label{tab:energy_windows}
\end{table}

For ${}^{225}$Ac, the cylinder of a Jaszczak phantom was used. Spheres of diameters $60~$mm, $37~$mm, and $28~$mm were filled at 1.37 kBq/mL and placed in the phantom with a 10:1 source to background ratio. Thirty-four SPECT acquisitions of the phantom were taken in sequence on a Siemens Symbia T2 system with the following settings: $128 \times 128$ pixels at $4.82~\text{mm} \times 4.82~\text{mm}$ resolution, 96 projection angles, high energy collimators, and 60~s acquisition time per projection with energy windows corresponding to gamma emission from ${}^{221}$Fr and ${}^{213}$Bi as indicated in Tab.\ \ref{tab:energy_windows}. To create clinically realistic scenarios, the projection angles were sub sampled to 32 angles post acquisition. In addition to reconstructing each photopeak separately, joint dual photopeak (JDP) reconstruction was also used, which used a stacked system matrix considering both peaks simultaneously (see H. Li in \cite{EANM2018}). In each case, the data were reconstructed using MLEM for up to 100 iterations. The VOIs masking the three spheres were considered for uncertainty estimation. Sample reconstructions, empirical uncertainties, and analytical uncertainties for these use cases are shown in Fig.\ \ref{fig:ac225_nema}.

\subsubsection{Dosimetry Simulation and Validation}

The following section considers propagation of uncertainty from VOIs (Eq.\ \ref{eq:cov_equation}) to TIA (Eq.\ \ref{eq:unc_cumulative_activity}) in both ${}^{177}$Lu and ${}^{225}$Ac imaging. The examples consist of using the anthropomorphic XCAT phantom \cite{XCat} in Monte Carlo simulations, where activity dynamics are defined to be analogous to the patient examples considered later. The true ``empirical'' variance $\tilde{u}(\hat{\mathscr{A}})$ obtained from TIA estimates of multiple SPECT noise realizations is compared to the estimated ``analytical'' variances $\hat{u}(\hat{\mathscr{A}})$ derived from single noise realizations.

An XCAT phantom was configured, analogous to the patient examples considered later, to have the following important regions: background, left kidney, right kidney, liver, a large, medium, and small liver lesion, bone lesions in the left and right hip bones. The organ biodistribution followed the mono-exponential curves in Tab.\ \ref{tab:fit_functions}, while the liver and bone lesions followed the bi-exponential curves in Tab.\ \ref{tab:fit_functions}. The activity concentration ratios between the ${}^{225}$Ac and ${}^{177}$Lu XCAT phantoms were 1:1000, corresponding to typical differences in injected activities between these isotopes.

Monte Carlo SPECT acquisition was performed in SIMIND \cite{simind}, where the scanner was representative of a typical Siemens Symbia T Series scanner. In particular, it contained a $9.5~$mm (3/8'') thick NaI scintillator crystal with $3.8~$mm FWHM intrinsic spatial resolution and $10\%$ energy resolution at $140~$keV (and $1/\sqrt{E}$ dependence). For ${}^{177}$Lu, a ``medium energy'' parallel hole collimator  with a hole length of $40.64~$mm and hole diameter of $2.94~$mm was used, and 96 projections were acquired for 15~s each. For ${}^{225}$Ac, a ``high energy'' parallel hole collimator with hole length of $59.70~$mm and hole diameter of $4.00~$mm was used, and 32 projections were acquired for 150~s each. Both simulations used a $0.4795~$cm pixel size with a projection matrix size of $128 \times 240$ in order to capture the full field of view, and a constant radial distance of $22~$cm. Acquired energy windows are shown in Tab.\ \ref{tab:energy_windows}. Noiseless projections were obtained and used to generate 100 independent realizations (${}^{177}$Lu) and 20 independent realizations (${}^{225}$Ac) for each time point of each isotope.

The time points considered for ${}^{177}$Lu were selected as 4~h, 28~h, 103~h, and 124~h to be consistent with subsequent patient examples; all 400 realizations were reconstructed using OSEM (4it8ss). The time points considered for ${}^{225}$Ac were 6~h, 21~h, 77~h, and 285~h to be consistent with subsequent patient examples. Reconstruction used a JDP system matrix; the upper peak employed a PSF model derived from SIMIND Monte Carlo data with the SPECTPSFToolbox \cite{spectpsftoolbox}, while the lower peak used standard Gaussian PSF modeling. All 80 realizations were reconstructed with MLEM (100it1ss + 3~cm Gaussian post filtering).  The images were scaled by (i) calibration factors (CFs) obtained via a point source simulation in SIMIND to convert counts to units of activity and (ii) by applying partial volume correction using measured recovery coefficients (RCs); the RCs were obtained by reconstructing corresponding noiseless projection data at each time point and computing the ratio of measured to true uptake in each VOI. Because many counts were simulated, the RCs and CFs had negligible uncertainties; this was done intentionally so that uncertainties in the TIA were purely due to Poisson noise in the projection data. Analytical uncertainties were estimated in all VOIs for each noise realization of each isotope.

For the following, let $j$ represent the time point index, $m$ represent the VOI index, and $i$ represent the noise realization index. Each set of four times points $\left\{t_j, A_j, \sigma_j \right\}_{i,m}$ were fit to the functions specified by Tab.\ \ref{tab:fit_functions} using the Trust Region Reflective algorithm available in the ``curve\_fit'' function of scipy. The curve fitting was performed three separate ways to obtain the estimates  $\hat{u}^{(\text{no-sig})}(\hat{\mathscr{A}}_{i,m}^{(\text{no-sig})})$, $\hat{u}^{(\text{prop-sig})}(\hat{\mathscr{A}}_{i,m}^{(\text{prop-sig})})$, and $\hat{u}^{(\text{est-sig})}(\hat{\mathscr{A}}_{i,m}^{(\text{est-sig})})$ respectively. The distribution of each these uncertainties (across $i$ and for each $m$) was compared to the corresponding empirical estimates $\tilde{u}^{(\text{no-sig})}(\hat{\mathscr{A}}_{m}^{(\text{no-sig})})$, $\tilde{u}^{(\text{prop-sig})}(\hat{\mathscr{A}}_{m}^{(\text{prop-sig})})$, and $\tilde{u}^{(\text{est-sig})}(\hat{\mathscr{A}}_{m}^{(\text{est-sig})})$ of the uncertainty obtained by computing the variance of each of the $A_{i,m}$ across the noise realizations $i$.

\subsubsection{Clinical Applications}

The uncertainty formalism was applied to two use cases on real patient data. In each case, Eq.\ \ref{eq:final} was used to obtain uncertainty estimates in VOIs at various time points, which were then used to compute the uncertainty on the TIA given by Eq.\ \ref{eq:unc_cumulative_activity}.

The first use case explored reconstruction of multi time point ${}^{177}$Lu-DOTATATE RPT data from the University of Michigan Deep Blue data sharing repository \cite{deepblue, dosimetry_part1, dosimetry_part2}. The patient considered was imaged on the day of injection, as well as 1, 4, and 5 days post-injection. Images were acquired on a Siemens Intevo system with a $128 \times 128$ matrix size and 120 projections with 25~s acquisition time per projection; acquired energy windows are shown in Tab.\ \ref{tab:energy_windows}. To study the effect of different numbers of acquired counts, the data were subsampled into the following three cases: (i) 120 projections at $25~$s / projection, (ii) 60 projections at $5~$s / projection, and (iii) 30 projections at $1.2~$s / projection. P data were reconstructed using OSEM (12it/8ss) for case (i), OSEM (24it/4ss) for case (ii), and OSEM (48it/1ss) for case (iii). In each case, five VOIs were considered for uncertainty analysis the left/right kidney and small/medium/large liver lesions; volumes were obtained by drawing VOIs that captured all counts from each lesion. Sample reconstructed MIPs and TACs are shown in Fig.\ \ref{fig:lu177_patient_recon_sample}, while explicit TIA and TIA uncertainty estimates for each case are shown in \ref{fig:lu177_patient_tia_stats}.

The second use case explored uncertainty estimation in bone metastasis of a patient receiving ${}^{225}$Ac-PSMA-617 therapy. The patient received an activity of 8~MBq and was imaged 6 h, 20.5 h, 76.5 h, and 284.6 h post injection on a GE Discovery 670 Pro SPECT/CT system. 30 projections (15 per head) were acquired for 150~s / projection; acquired energy windows are shown in Tab.\ \ref{tab:energy_windows}. A GE high energy general purpose collimator was used for image acquisition. 

Reconstruction used a JDP system matrix; the upper peak employed a PSF model derived from SIMIND Monte Carlo data with the SPECTPSFToolbox \cite{polson2024fastaccuratecollimatordetectorresponse}, while the lower peak used standard Gaussian PSF modeling. To account for stray radiation-related noise due to the low count rate, a blank scan was taken using the same acquisition parameters, and the mean background counts in each energy window was taken into consideration in reconstruction. Images at each time point were reconstructed via MLEM (100it) with a 3~cm post-reconstruction filter using (i) the 218~keV peak, (ii) the 440~keV peak, and (iii) using both peaks simultaneously. For (iii), the relative primary rates between the peaks was estimated by simulating a point source with a similar imaging system in SIMIND. 

Three lesions were segmented by a physician on a pre-treatment PET image using the PET Edge+ tool of MIM v7.2.1 (MIM Software Inc., USA), and were used as VOI masks for estimating uncertainty via Eq.\ \ref{eq:final}. The four time points were fitted to the lesion activity fit function in Tab.\ \ref{tab:fit_functions}, and the uncertainty on the TIA was estimated using Eq.\ \ref{eq:unc_cumulative_activity} with $\hat{V}^{(\text{est-sig})}$.

\section{Results}

\subsection{Uncertainty Estimation Validation}

\begin{figure}[h]
\centering
\includegraphics[width=\columnwidth]{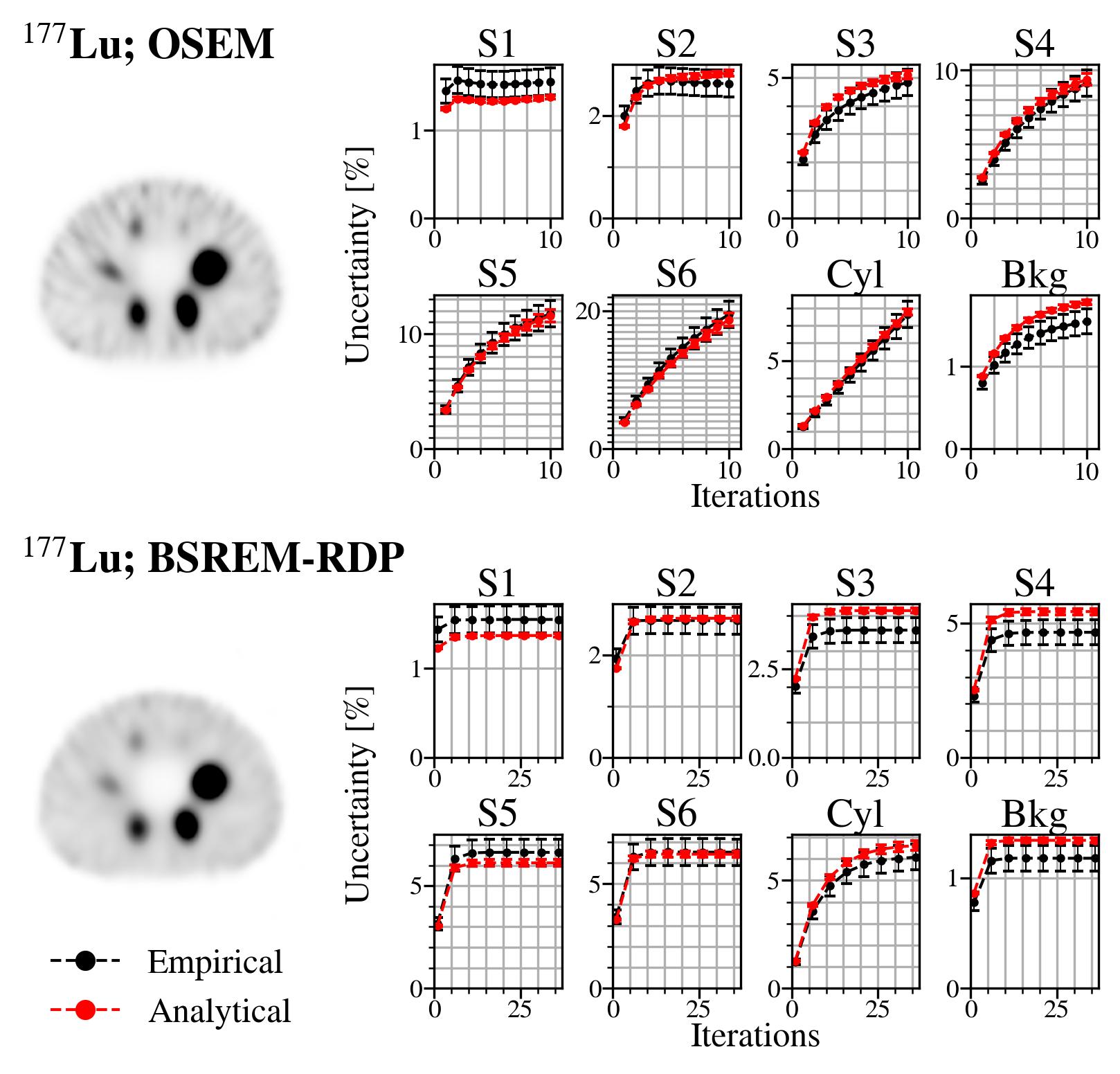}
\caption{Reconstruction and uncertainty estimation of a ${}^{177}$Lu NEMA phantom acquired on a Siemens Symbia T2 scanner. Left: axial slices of the reconstructed NEMA phantom from a sample acquisition corresponding to 4 iterations (OSEM) and 40 iterations (BSREM). Right: uncertainty estimates for eight VOIs as a function of iteration number; shown in black is the empirical uncertainty $\tilde{\delta}(\hat{x}^k \cdot \xi)$ obtained from all noise realizations via Eq.\ \ref{eq:rel_unc_emp} (error bars correspond to the standard error of the standard deviation estimator); shown in red is mean and standard deviation (error bars) of the analytically obtained uncertainties $\hat{\delta}(\hat{x}^k \cdot \xi)$ from Eq.\ \ref{eq:final}}.
\label{fig:lu177_nema}
\end{figure}

The results of the ${}^{177}$Lu phantom study are shown in Fig.\ \ref{fig:lu177_nema}. In all regions, the computed analytical uncertainty were consistent with the empirically obtained uncertainty estimates, providing support for Eq.\ \ref{eq:final} in this activity regime. For OSEM, the uncertainties increased with more iterations and smaller volumes, while for BSREM-RDP, the uncertainties tended to converge in all regions after a few iterations, with smaller regions yielding larger uncertainties after convergence. The uncertainties in BSREM were smaller than OSEM for the small spheres after many iterations.

\begin{figure}[h]
\centering
\includegraphics[width=\columnwidth]{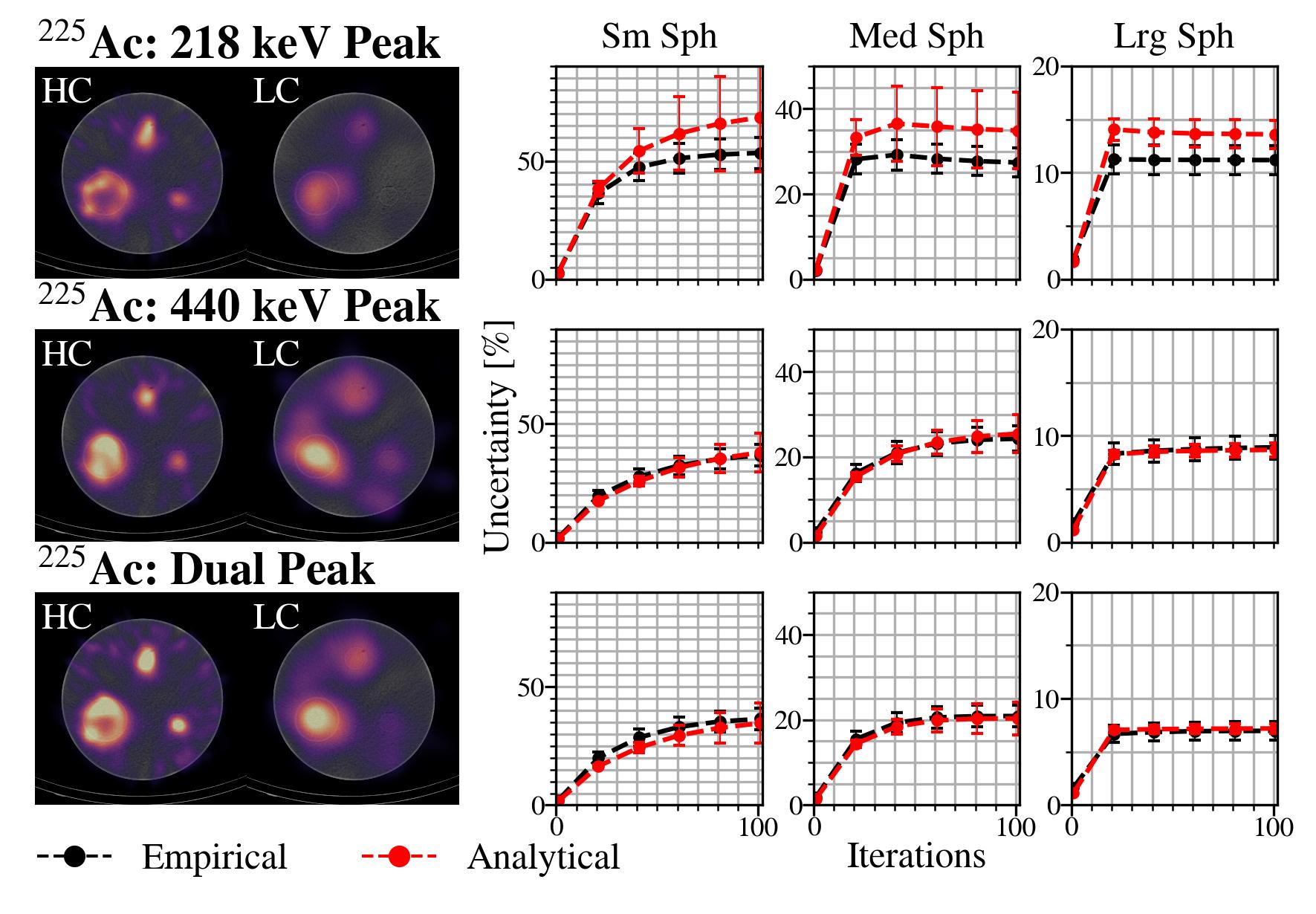}
\caption{Reconstruction and error estimation for the ${}^{225}$Ac NEMA phantom. From top to bottom are reconstructions using 218~keV peak, 440~keV peak, and the dual peaks. High count (HC) was obtained by adding all 33 sets of projection data together and low count (LC) was a reconstruction of a sample acquired image. Images correspond to axial slices of reconstruction (MLEM 100it) with 3~cm Gaussian post filtering used in the LC case. Shown in black is the empirical uncertainty obtained via Eq.\ \ref{eq:rel_unc_emp} with error bars corresponding to the standard error of this statistic. Shown in red is the analytically obtained uncertainty from Eq.\ \ref{eq:final}, with error bars corresponding to the standard deviation across estimates for each acquisition.}\label{fig:ac225_nema}
\end{figure}

The results of the ${}^{225}$Ac phantom study are shown in Fig.\ \ref{fig:ac225_nema}. For reconstruction using the $440~$keV peak and the dual peak, computed analytical uncertainties had little variance between different acquisitions, and their estimates were consistent with the empirically estimate uncertainties. Reconstruction using the 218~keV yielded analytical uncertainties with significant variance in the small ($\pm 22.8\%$) and medium ($ \pm 8.9\%$) spheres, likely due to the smaller count rate in this energy window.

\subsection{Dosimetry Simulation and Validation}
\begin{figure}[h]
\centering
\includegraphics[width=\columnwidth]{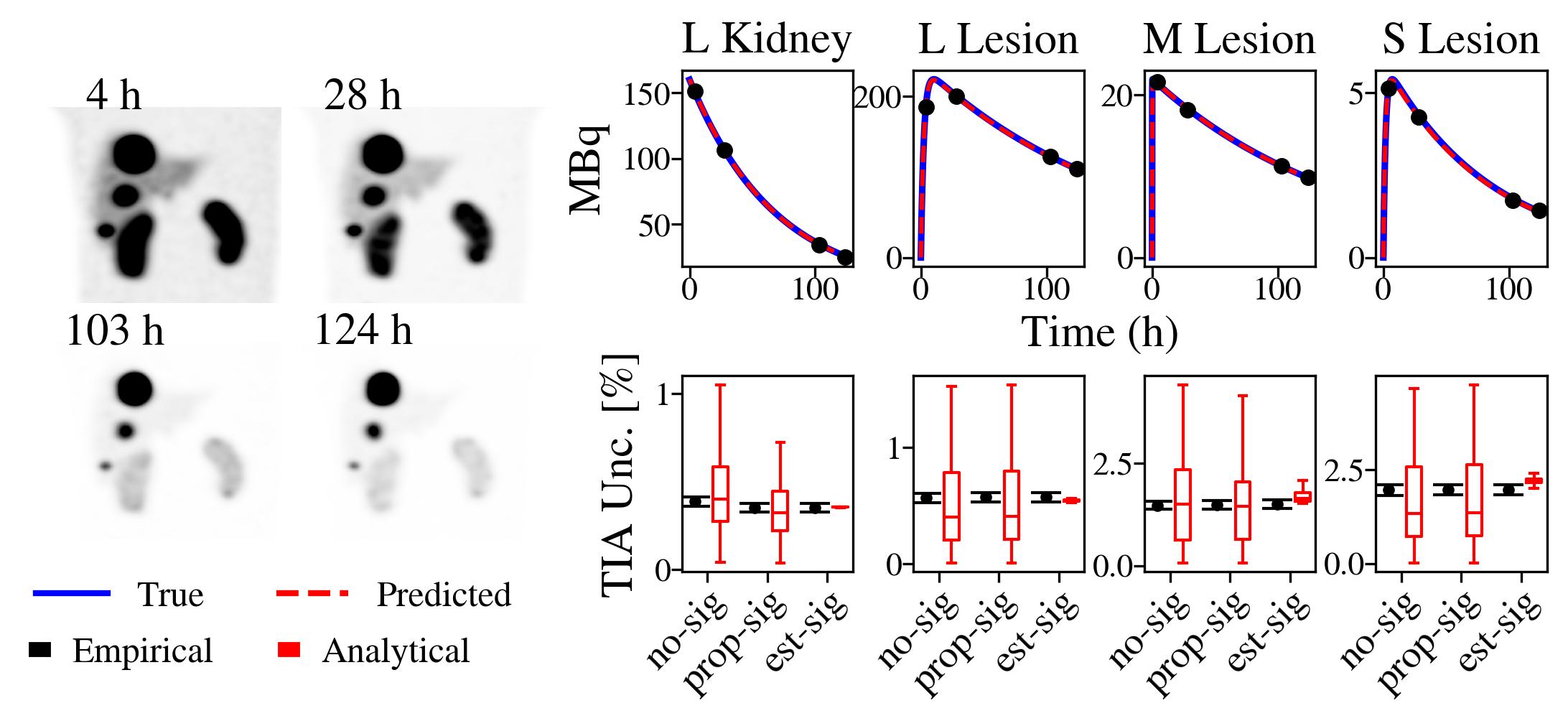}
\caption{Results from ${}^{177}$Lu XCAT simulations. Left: sample maximum intensity projections of reconstructions at each of the 4 time points (OSEM 4it8ss). Right: Quantities of interest from simulations where each column in the plot array corresponds to a particular VOI. Top right: true TAC (blue), estimated TAC (dotted red) and data points (black). Bottom right: Box plots of predicted TIA uncertainties $\hat{u}(\hat{\mathscr{A}})$ histogrammed over the 100 trials (red) and empirical uncertainties of TIA $\tilde{u}(\hat{\mathscr{A}})$ evaluated by directly computing the variance in $\hat{\mathscr{A}}$ over multiple trials (black); the box extends from the first to the third quartile with a line at the median, and the whiskers extend to the farthest data point lying within the inter quartile range. }\label{fig:xcat_sims_lu177}
\end{figure}

\begin{figure}[h]
\centering
\includegraphics[width=\columnwidth]{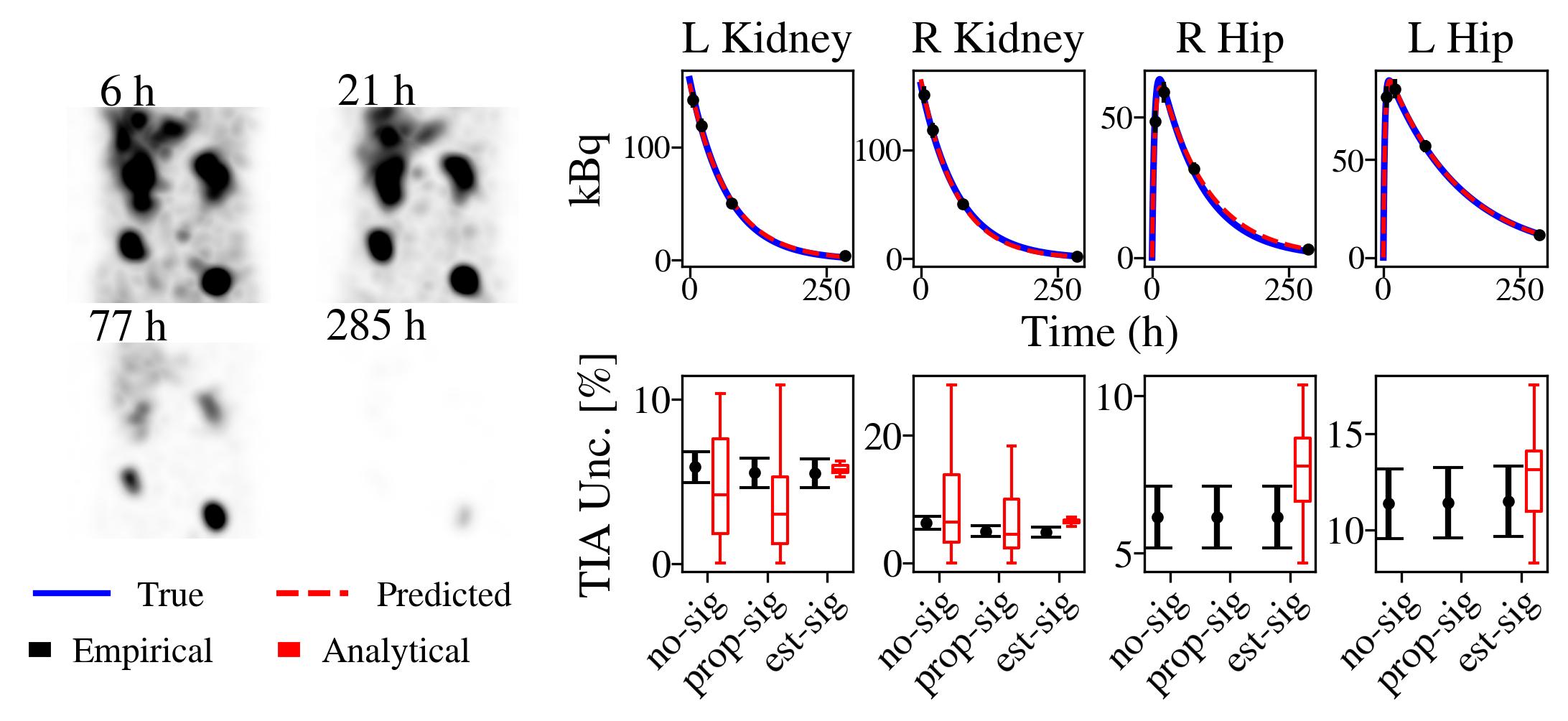}
\caption{Results from ${}^{225}$Ac XCAT simulations. Left: Sample coronal slices of reconstructions at each of the 4 time points (MLEM 100it+3~cm Gaussian smoothing). Right: Quantities of interest from simulations where each column in the plot array corresponds to a particular VOI. Top right: true TACs (blue), sample fitted TACs (dotted red) and data points (black). Bottom right: Box plots of analytical uncertainties in TIA $\hat{u}(\hat{\mathscr{A}})$ histogrammed over the 20 trials (black) and empirical uncertainties of TIA $\tilde{u}(\hat{\mathscr{A}})$ evaluated by directly computing the variance in $\hat{\mathscr{A}}$ over multiple trials (red); the box extends from the first to the third quartile with a line at the median, and the whiskers extend to the farthest data point lying within the inter quartile range. No-sig and prop-sig covariance matrices could not be estimated for the lesions in the left/right hip: uncertainties are thus not shown.}\label{fig:xcat_sims_ac225}
\end{figure}

Figs.\ \ref{fig:xcat_sims_lu177} and \ref{fig:xcat_sims_ac225} show the results of the Monte Carlo dosimetry validation study. For all VOIs considered, use of the uncertainties given by Eq.\ \ref{eq:final} in TAC fitting yielded accurate TIA uncertainty predictions with significantly improved precision relative to use of (i) proportional uncertainties given by Eq.\ \ref{eq:rel2abs_V} and (ii) no uncertainties. In both isotopes, the uncertainty on the TIA was larger in regions modeled by a bi-exponential. For example, although the large lesion had greater activity than the left kidney at every time point in the ${}^{177}$Lu simulation, it also had a larger estimated TIA uncertainty of $(0.817 \pm 0.014)\%$ (vs. $(0.404 \pm 0.001)\%)$. In the ${}^{225}$Ac case, the covariance on the parameters for the right and left hip bone lesions could not be estimated using no and proportional uncertainties; this is elaborated upon in the discussion section.

\subsection{Clinical Applications}

\begin{figure}[h]
\centering
\includegraphics[width=\columnwidth]{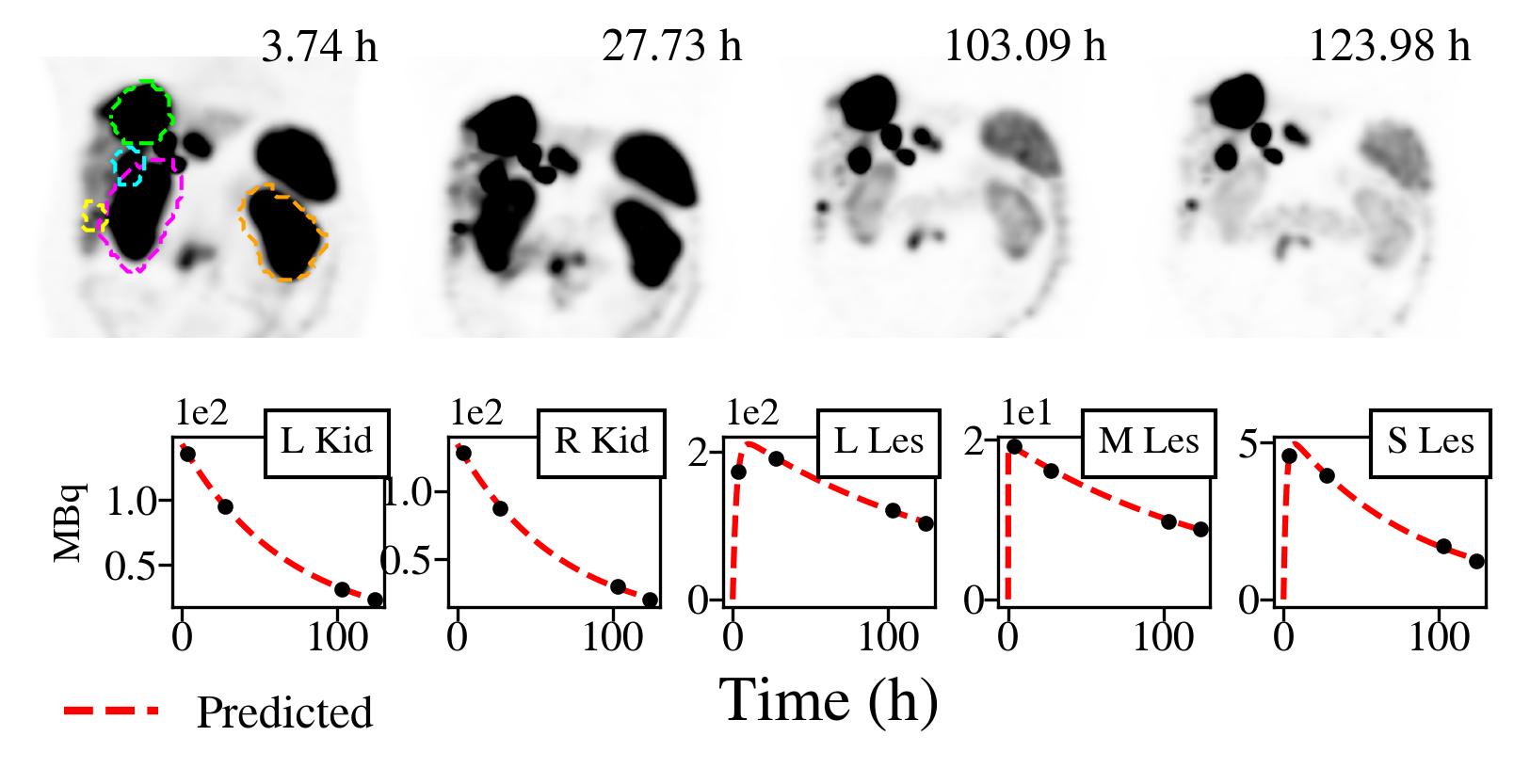}
\caption{Reconstructions and TACs of ${}^{177}$Lu-DOTATATE data from the SNMMI Dosimetry Challenge (25~s / projection). Top: maximum intensity projections with the top left image showing the segmentations of the left kidney (orange), right kidney (pink), large lesion (green), medium lesion (blue), and small lesion (yellow). Bottom: predicted activity at each time point (black) and TACs (red).}\label{fig:lu177_patient_recon_sample}
\end{figure}

\begin{figure}[h]
\centering
\includegraphics[width=\columnwidth]{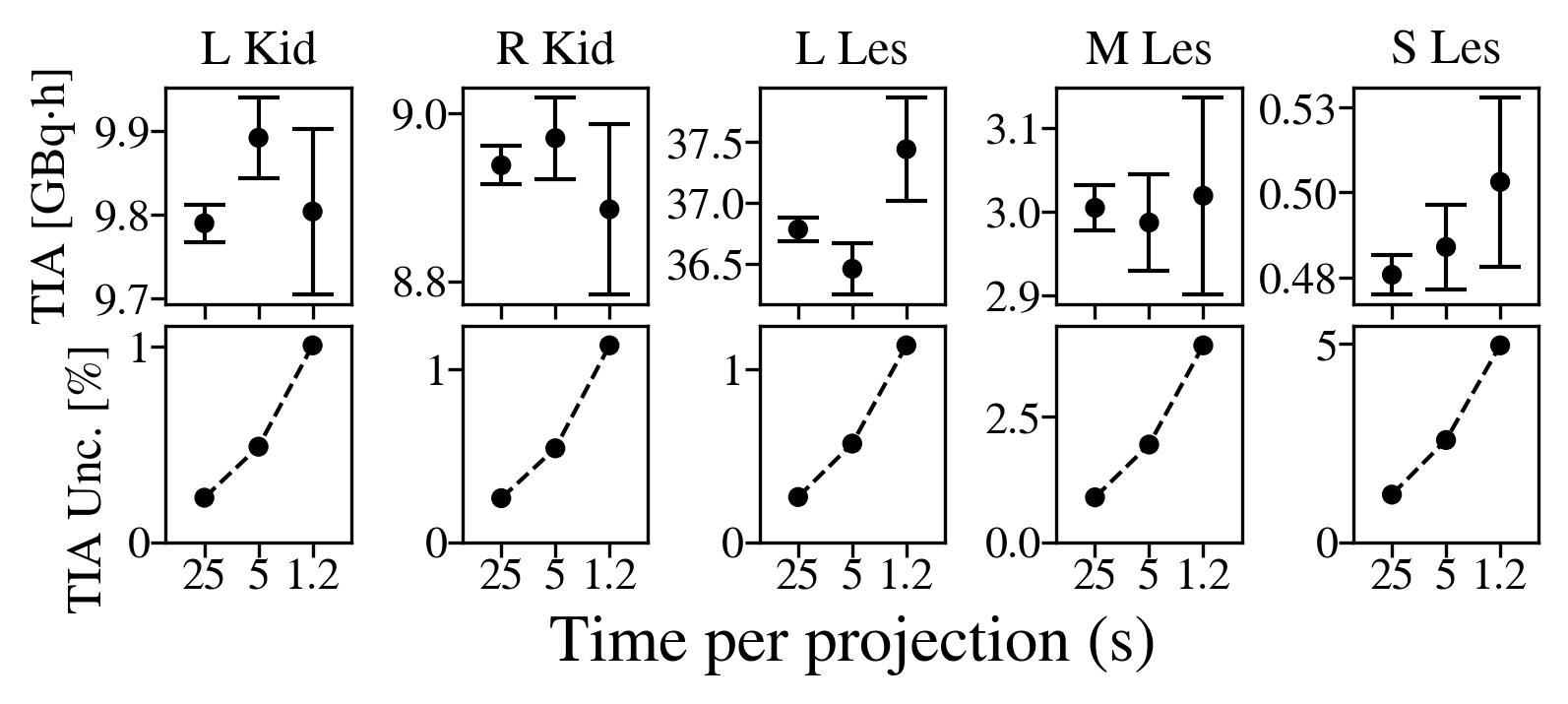}
\caption{Predicted TIA and uncertainties (top) and uncertainty only (bottom) in left kidney (L Kid), right kidney (R Kid) large lesion (L Les) medium lesion (M Les) and small lesion (S Les) for the three different subsampled cases of the SNMMI dosimetry challenge data: (i) 25~s / projection, (ii) 5~s / projection and (iii) 1.2~s / projection.}\label{fig:lu177_patient_tia_stats}
\end{figure}

Sample reconstructions and TACs for the ${}^{177}$Lu case with 120 projections and 25 s / projection are shown in Fig.\ \ref{fig:lu177_patient_recon_sample}. TIA and TIA uncertainty plots for all three subsampled acquisition times are shown in Fig.\ \ref{fig:lu177_patient_tia_stats}. The uncertainties in the kidney TIAs were small ($<3\%$), even when the projection time is reduced from 25~s / projection to 1.2~s / projection. The uncertainty in the three liver lesions increased more substantially as the time per projection was reduced, increasing from $1.26\%$ (25~s / projection) to $5.11\%$ (1.2~s / projection) in the smallest lesion. In general, the TIA error became larger with smaller volumes. Despite the large lesion having a TIA three times greater than the left kidney, its TIA uncertainty was similar, likely due to the larger number of TAC parameters (2 for mono-exponential vs. 3 for bi-exponential).

\begin{figure}[h]
\centering
\includegraphics[width=\columnwidth]{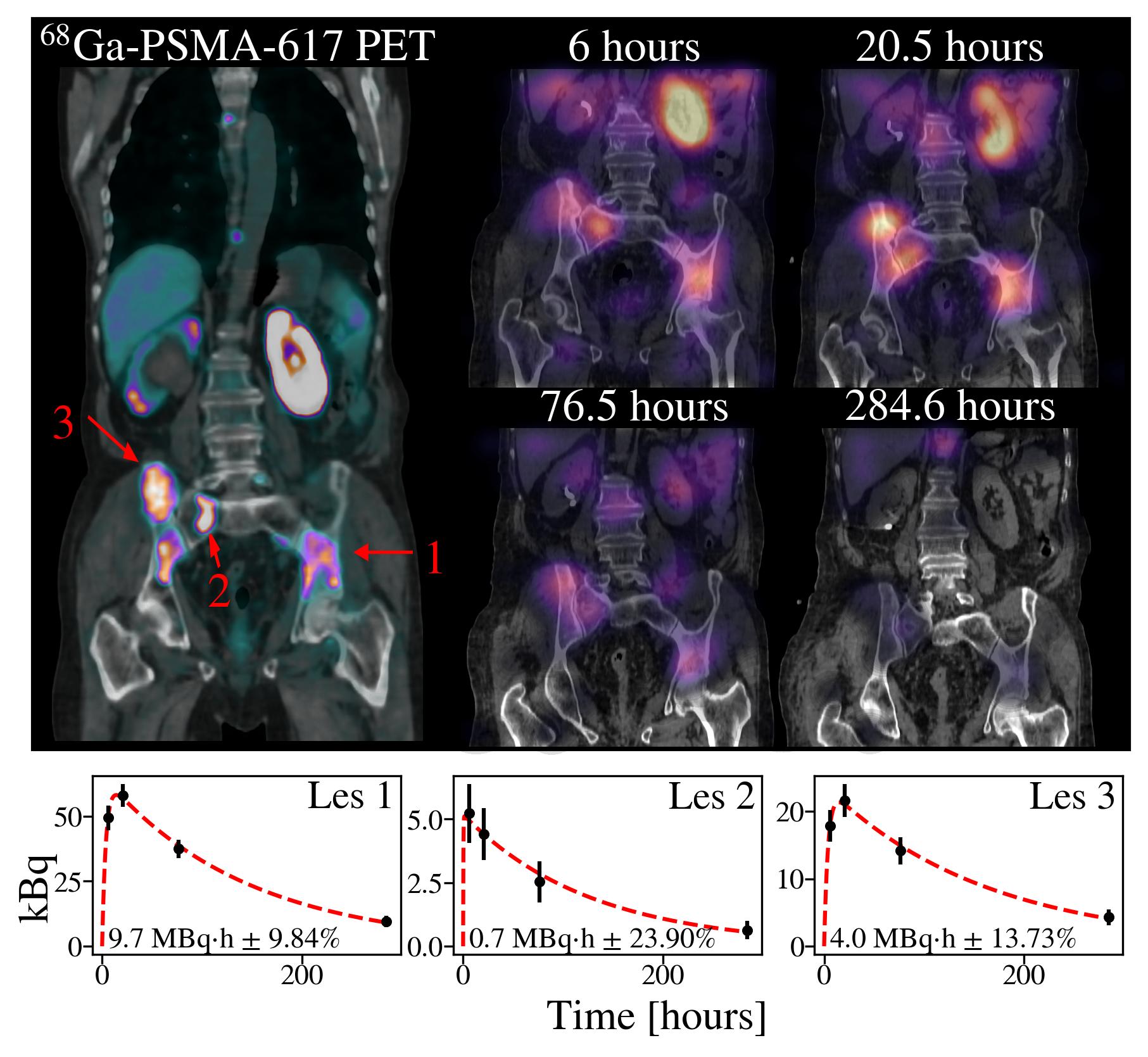}
\caption{Reconstruction, TACs, and TIA estimation of multi time point ${}^{225}$Ac-PSMA-617 data using JDP in reconstruction. Top: coronal slices of a  pre therapy PET image and reconstructed SPECT images at each time point. Lesions numbers are marked in red (lesion 2 is directly above the annotation). Bottom: TACs for each lesion; TIA and TIA uncertainty is printed on each plot.}\label{fig:ac225_patient_recon_sample}
\end{figure}

\begin{figure}[h]
\centering
\includegraphics[width=\columnwidth]{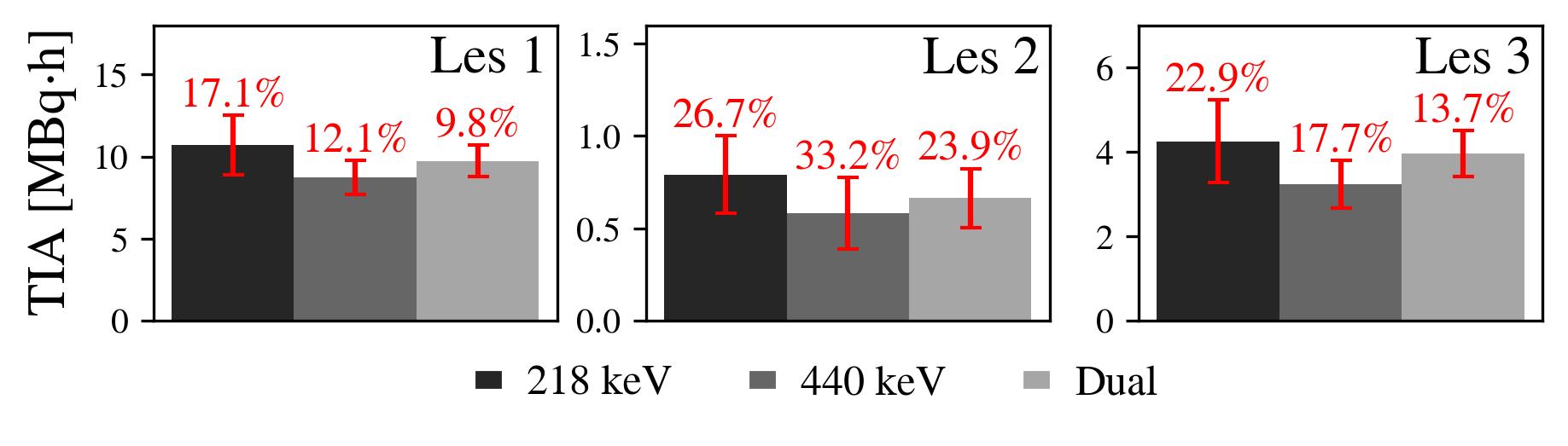}
\caption{Predicted TIAs (bars) and uncertainties (red) for the ${}^{225}$Ac-PSMA-617 data of Fig.\ \ref{fig:ac225_patient_recon_sample} in pevlis and sacrum via reconstruction with (i) the 218~keV photopeak, (ii) the 440~keV photopeak and (iii) dual photopeak, which uses both peaks simultaneously.}\label{fig:ac225_patient_tia_stats}
\end{figure}

Sample reconstructions and TACs for the ${}^{225}$Ac case with dual peak reconstruction are shown in Fig.\ \ref{fig:ac225_patient_recon_sample}. TIA and TIA uncertainty plots for lesions in the pelvis and sacrum are shown in Fig.\ \ref{fig:ac225_patient_tia_stats}. Though the dual peak TIA uncertainties were always smallest, the estimated uncertainties of  $9.84\%$, $23.90\%$, and $13.73\%$ in the three lesions were significantly larger than in the ${}^{177}$Lu example.

\section{Discussion}
The VOI-based uncertainty technique of Eq.\ \ref{eq:final} was tested on ${}^{177}$Lu and ${}^{225}$Ac SPECT phantom data (and ${}^{18}$F PET phantom data in the appendix) by comparing uncertainty estimations from single acquisitions to empirical uncertainty estimates given by the variation across multiple acquisitions. The method was validated for both the OSEM and BSREM-RDP reconstruction algorithms. It was shown consistent with the empirical estimates in nearly all regions, with the exception of those with very low activity. Estimated uncertainties from Eq.\ \ref{eq:final} were then used in TAC fitting from Monte Carlo SPECT simulations. It was demonstrated that their use yielded accurate TIA uncertainty estimates with far greater precision than (i) use of no uncertainty and (ii) use of proportional uncertainties of VOI activity. For ${}^{177}$Lu, the TIA uncertainty estimates had better precision in regions with biodistributions modeled by mono-exponential biokinetics (compared to bi-exponential) due to the smaller number of model parameters in curve fitting. For ${}^{225}$Ac, TIA uncertainties could only be estimated for bone lesions if Eq.\ \ref{eq:final} was used; since there were only slightly more data points than curve fit parameters, the under-constrained problem yielded near zero values for $\chi^2$ and thus unstable estimates in Eq.\ \ref{eq:rel2abs_V}. This further highlights the importance of the proposed method, since there may be cases where TIA uncertainty cannot be estimated unless the LSC uncertainty is known at each time point. Finally, the method was applied to real clinical data to demonstrate the range of expected TIA uncertainties in ${}^{177}$Lu-DOTATATE and ${}^{225}$Ac-PSMA-617 imaging.

There are a few noteworthy observations from the validation section. The predicted analytical uncertainties were generally consistent with the empirical uncertainties, except in regions with low activity such as (i) the small and medium spheres for ${}^{225}$Ac (218 keV peak) and (ii) the cold cylinder for ${}^{18}$F from Fig.\ \ref{fig:f18_nema} in the appendix. In these regions, the variance of the analytical uncertainties was large, highlighting that the method failed for certain noise realizations. In Fig.\ \ref{fig:ac225_nema}, the displayed ${}^{225}$Ac low count 218~keV peak image (which corresponds to a particular noise realization) had almost no activity predicted in the small sphere: this impacted the uncertainty estimation algorithm. These results do not invalidate the theory, however, since they are evidently scenarios where use of $\hat{x}^k$ as a proxy for $\bar{\hat{x}}^k$ in Eq.\ (\ref{eq:cov_equation}) fails. These extreme cases demonstrate the limitations of the derived uncertainty technique. For ${}^{177}$Lu and ${}^{18}$F, OSEM resulted in higher VOI uncertainties than BSREM-RDP. Regularized algorithms, however, resulted in VOIs with smaller RCs and greater activity leakage from outside regions. These larger systematic errors need to be considered against the smaller LSC uncertainties.

A minor limitation of the SPECT phantom experiments is that the activity of ${}^{177}$Lu and ${}^{225}$Ac decreased throughout the repeated acquisitions that each spanned over 24 hours. While the counts in each VOI were scaled by an exponential scalar weighting in Eq.\ \ref{eq:rel_unc_emp} to remove the component of variation resulting from activity decay, it was assumed that the relative uncertainties in each VOI remained approximately constant over time. In practice, the uncertainty at later time points may be slightly higher due to the reduced number of counts measured.

The curve fitting validation also has limitations that may not have been addressed here. In the XCAT simulation, the time dependent distribution of activity truly followed mono-exponential and bi-exponential distributions before simulation of the SPECT acquisition. In a real clinical scenario, there may be sources of uncertainty in the true organ activity, perhaps related to stochastic biological phenomena. These additional sources of uncertainty, which are separate from the imaging system, would need to be accounted for when estimating the covariance matrix of the TAC parameters. Future research might aim to consider these sources of uncertainty in Monte Carlo simulations and determine the importance of including these uncertainties in TAC fitting.

While uncertainty propagation in scaling the TIA by the system CFs is straightforward since the uncertainties can be added in quadrature after TAC fitting, propagation of uncertainty in RCs may be more nuanced, since each time point may have a separate RC dependent on the surrounding activity. In this case, each individual time point would need to be scaled by a separate RC before curve fitting, and the uncertainty would need to be included in the curve fitting procedure. If the same RC is used for every time point, however, then recovery correction can be applied directly to the TIA after TAC fitting. 

Another potential issue is correlation between VOI segmentations and estimated uncertainties from SPECT images. If VOIs are segmented solely using external images, such as CT or PET, then no such correlation would exist. However, if activity distributions were used to guide segmentations, then SPECT noise realizations (and any associated estimated uncertainty from Eq.\ \ref{eq:final}) are correlated with VOI segmentation; this would need to be accounted for in dose uncertainty protocols. While consideration of these factors is beyond the scope of this paper, they are of high priority in future studies.

The application of Eq.\ \ref{eq:final} for real data TIA uncertainty estimation was meant to demonstrate the feasibility of this technique in dosimetry use cases. In the ${}^{177}$Lu-DOTATATE example, all VOI uncertainties for the standard imaging protocol (25~s / projection) are small relative to other uncertainties in the dosimetry pipeline, such as those propagating from RCs and CFs ($8.2$\% (medium lesion) and $6.6$\% (Symbia T2) respectively in Nuttens et.\ al.\ \cite{nuttens_uncertainty}). For the ${}^{225}$Ac-PSMA-617 RPT use case, it is demonstrated in Fig.\ \ref{fig:ac225_patient_tia_stats} that use of dual peak reconstruction yields the lowest uncertainty in the TIA. It should be emphasized that the TIA estimates in Fig.\ \ref{fig:ac225_patient_tia_stats} still need to be adjusted based on RCs, which may potentially be different for the 218~keV, 440~keV, and dual energy window reconstructions separately. Since the estimated uncertainties of $9.84\%$, $23.90\%$, and $13.73\%$ for the three lesions are substantial compared to RC and CF uncertainties from Nuttens et.\ al.\ \cite{nuttens_uncertainty}, it follows that the uncertainty estimation technique derived here is essential to account for all sources of uncertainty in ${}^{225}$Ac dosimetry.

Future studies should aim to study the impact of including the proposed uncertainty method in a full dosimetry protocol. Furthermore, although the TIA uncertainties may be relatively insignificant for standard imaging times for ${}^{177}$Lu RPTs, they could be used to find patient specific minimum scan time requirements. For example, data from the first time point of an RPT procedure can be sub-sampled and used to establish a relationship between computed uncertainty and projection time in relevant VOIs, and used to create a minimum bound on the required acquisition time for the patient in subsequent scans.

\section{Conclusion}
In emission tomography, uncertainty propagates from the Poisson-distributed random counts in the projection data to reconstructed images. This work proposed a computationally efficient algorithm to estimate the uncertainty on measured counts from volumes of interest in reconstructed images.  The algorithm was validated by acquiring repeated ${}^{177}$Lu and ${}^{225}$Ac SPECT scans of a physical phantom, and demonstrating that the variability of total counts in select volumes of interest across multiple acquisitions was consistent with the estimated uncertainties obtained from a single acquisition. Monte Carlo simulations demonstrated that use of the estimated uncertainties in time activity curve fitting and time integrated activity estimation yield uncertainty estimates that are more accurate and precise than conventional methods that use uncertainties proportional to square root of counts. For ${}^{225}$Ac-PSMA-617 patient data, application of the algorithm demonstrated that uncertainty in the projection data contributes significantly to uncertainty on time integrated activity, and that this uncertainty is large relative to other uncertainties attained in dosimetry. In summary, this work has provided (i) validation of proposed uncertainty algorithm and (ii) a blueprint for how the algorithm can be applied in dosimetry protocols. The algorithm has been implemented in the open source python library PyTomography to maximize outreach to the nuclear medicine community.

\appendix

 \subsection{Regional Count Variability}

\begin{figure}[h]
\centering
\includegraphics[width=\columnwidth]{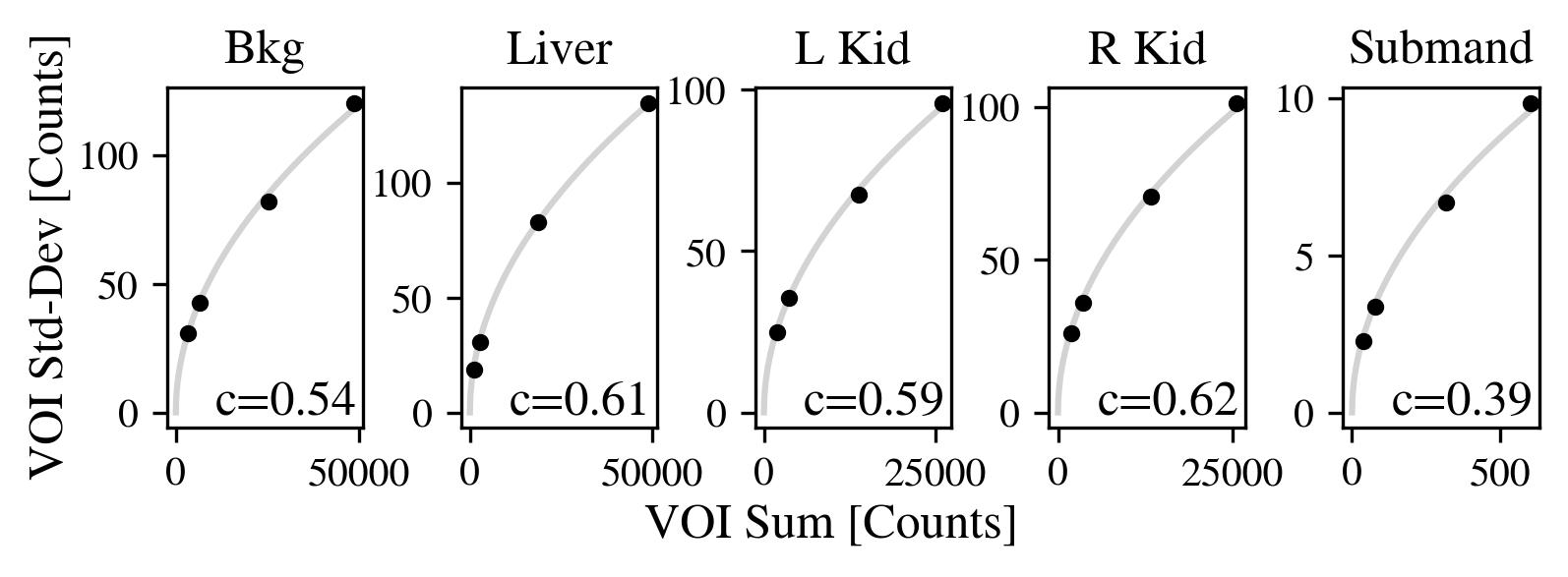}
\caption{Standard deviation of VOI counts across noise realizations vs. VOI counts, in five different organs from  ${}^{177}$Lu-PSMA-617 XCAT simulations.}\label{fig:appendix}
\end{figure}

To justify the claim made by Eq.\ \ref{eq:prop_sigma}, the count standard deviation in the kidneys and liver lesions for separate noise realizations of the ${}^{177}$Lu-PSMA-617 XCAT simulation is plotted as a function of the total VOI counts in Fig.\ \ref{fig:appendix}. Each set of data were fit to the functional form $\sigma = c\sqrt{\hat{x} \cdot \xi}$ where $c$ is a proportionality factor. As demonstrated, the proportionality exists, but is different for each VOI.

 \subsection{PET Application}
\begin{figure}[h]
\centering
\includegraphics[width=\columnwidth]{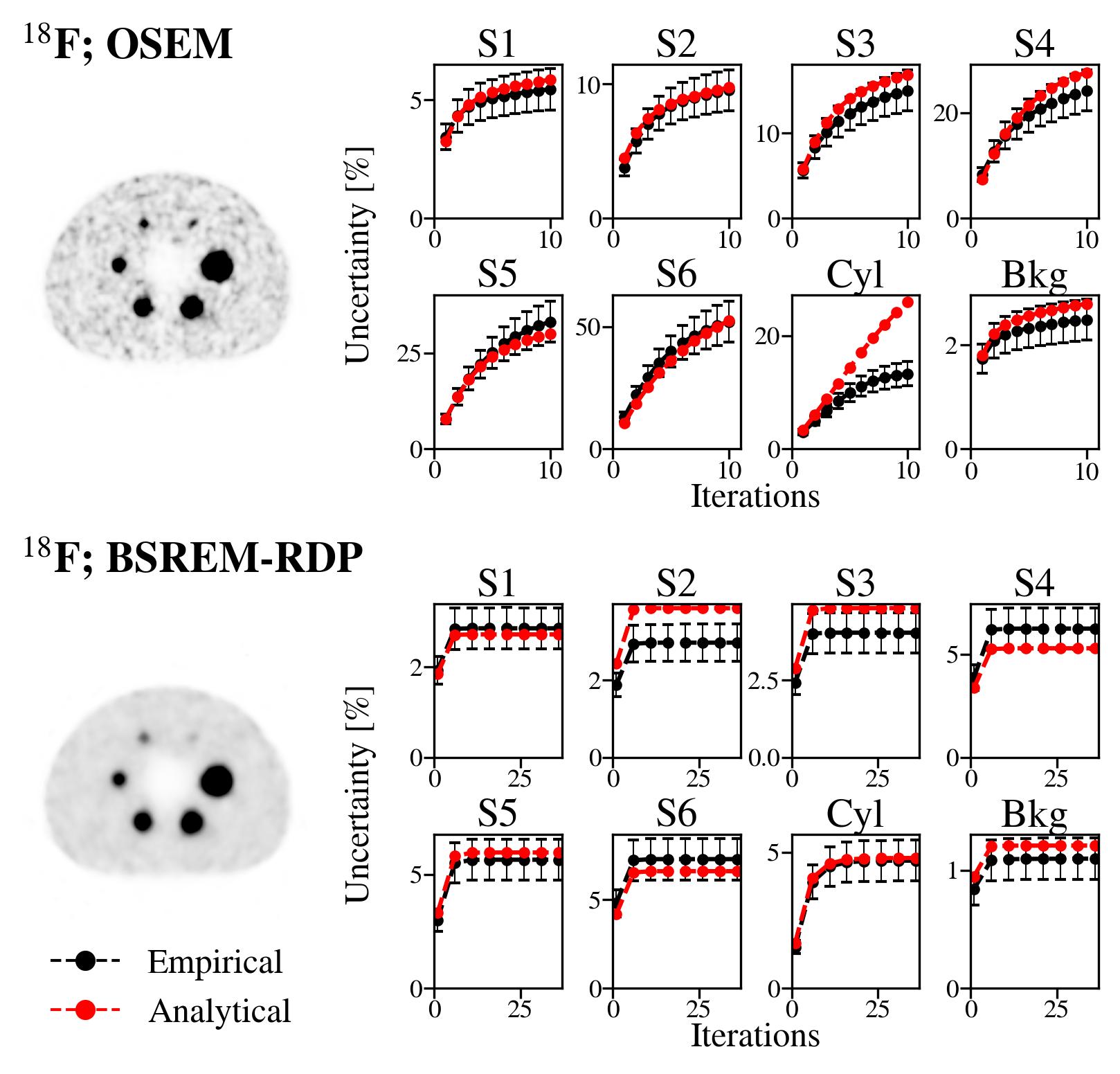}
\caption{Reconstruction and error estimation for the ${}^{18}$F NEMA phantom acquired on a GE Discovery MI scanner. Left: axial slices of the reconstructed NEMA phantom from a sample acquisition. Right: uncertainty estimates for eight VOIs as a function of iteration number; shown in black is the empirical uncertainty $\tilde{\delta}(\hat{x}^k \cdot \xi)$ obtained via Eq.\ \ref{eq:rel_unc_emp} (error bars correspond to the standard error of the standard deviation estimator); shown in red is the proposed analytical uncertainty estimate $\hat{\delta}(\hat{x}^k \cdot \xi)$ from Eq.\ \ref{eq:final}. Error bars are not visible for the analytically obtained uncertainty due to their small magnitude.}\label{fig:f18_nema}
\end{figure}

The uncertainty method was also validated for PET reconstructions using publicly available ${}^{18}$F NEMA phantom data \cite{parallelproj} acquired on a GE Discovery MI scanner (this is a proof-of-concept assessment, and can be extended to different applications in PET imaging such as ${}^{90}$Y dosimetry ). The acquired listmode data were split into 20 equally sized sets; each subset was reconstructed, and uncertainties were estimated for each of the six NEMA spheres, a VOI in the background, and the central cold cylinder portion of the phantom. Scatter and normalization for each list mode event was obtained using the GE vendor software Duetto, while images were reconstructed and errors estimated using PyTomography. Sample reconstructions, analytical uncertainties, and estimated uncertainties for these use cases are shown in Fig.\ \ref{fig:f18_nema}. Results are elaborated in the discussion.

\bibliographystyle{IEEEtran}
\bibliography{text}

\end{document}